\documentclass[bibyear]{aa} 

\usepackage{graphicx}
\usepackage{newclude}
\usepackage{url}
\usepackage{txfonts}
\usepackage{cancel}
\usepackage{natbib}
\bibpunct{(}{)}{;}{a}{}{,} 
\begin{document}
   \title{Prompt emission of GRB 121217A from $\gamma$-rays to the NIR}
   \author{J.~Elliott\inst{1} \and 
           H.~-F.~Yu\inst{1} \and
           S.~Schmidl\inst{2} \and
           J.~Greiner\inst{1} \and
           D.~Gruber\inst{1} \and
           S.~Oates\inst{3} \and
           S.~Kobayashi\inst{4} \and
           B.~Zhang\inst{5} \and
           J.~R.~Cummings\inst{6,7} \and
           R.~Filgas\inst{8} \and
           N.~Gehrels\inst{9} \and
           D.~Grupe\inst{10} \and
           D.~A.~Kann\inst{1} \and
           S.~Klose\inst{2} \and
           T.~Kr{\"u}hler \inst{11,12} \and
           A.~Nicuesa~Guelbenzu\inst{2} \and
           A.~Rau\inst{1} \and
           A.~Rossi\inst{2} \and
           M.~Siegel\inst{10} \and
           P.~Schady\inst{1} \and
           V.~Sudilovsky\inst{1} \and
           M.~Tanga\inst{1} \and
           K.~Varela\inst{1}
           }

   \institute{
	  Max-Planck-Institut f{\"u}r extraterrestrische Physik, Giessenbachstra{\ss}e 1, 85748, Garching, Germany.\\
              \email{jonnyelliott@mpe.mpg.de} \and 
          Th\"{u}ringer Landessternwarte Tautenburg, Sternwarte 5, 07778, Tautenburg, Germany. \and
          Mullard Space Science Laboratory, University College London, Holmbury St. Mary, Dorking Surrey, RH5 6NT. \and
          Astrophysics Research Institute, Liverpool John Moores University, IC2, Liverpool Science Park, 146 Brownlow Hill, Liverpool L3 5RF, United Kingdom. \and 
          Department of Physics and Astronomy, University of Nevada, Las Vegas, NV89154, USA. \and 
          University of Maryland Baltimore County, 1000 Hilltop Circle, Baltimore, MD 21250, USA. \and 
          Center for Research and Exploration in Space Sciences and Technology, NASA Goddard Space Flight Center, Greenbelt, MD 20771, USA. \and 
	  Institute of Experimental and Applied Physics, Czech Technical University in Prague, Horsk{\'a} 3a/22, 12800, Prague, Czech Republic. \and 
          Astrophysics Science Division, NASA, Goddard Space Flight Center, Greenbelt, MD 20771, USA. \and 
          Department of Astronomy and Astrophysics, Pennsylvania State University, 525 Davey Laboratory, University Park, PA 16802, USA. \and 
          Dark Cosmology Centre, Niels Bohr Institute, University of Copenhagen, Juliane Maries Vej 30, 2100 K\o benhavn \O, Denmark. \and
          European Southern Observatory, Alonso de C\'{o}rdova 3107, Vitacura, Casilla 19001, Santiago 19, Chile.
           }
   \date{}

  \abstract{The mechanism that causes the prompt-emission episode of $\gamma$-ray bursts (GRBs) is still widely debated despite there being thousands of prompt detections. The favoured internal shock model relates this emission to synchrotron radiation. However, it does not always explain the spectral indices of the shape of the spectrum, often fit with empirical functions, such as the Band function. Multi-wavelength observations are therefore required to help investigate the possible underlying mechanisms that causes the prompt emission. We present GRB 121217A, for which we were able to observe its near-infrared (NIR) emission during a secondary prompt-emission episode with the Gamma-Ray Burst Optical Near-infrared Detector (GROND) in combination with the {\it Swift} and {\it Fermi} satellites, covering an energy range of 5 orders of magnitude ($10^{-3}\, \rm keV$ to $100\, \rm keV$). We determine a photometric redshift of $z=3.1\pm0.1$ with a line-of-sight with little or no extinction ($A_{V}\sim0\, \rm mag$) utilising the optical/NIR SED. From the afterglow, we determine a bulk Lorentz factor of $\Gamma_{0}\sim250$ and an emission radius of $R<10^{18}\, \rm cm$. The prompt-emission broadband spectral energy distribution is well fit with a broken power law with $\beta_{1}=-0.3\pm0.1$, $\beta_{2}=0.6\pm0.1$ that has a break at $E=6.6\pm0.9\, \rm keV$, which can be interpreted as the maximum injection frequency. Self-absorption by the electron population below energies of $E_{a}<6\, \rm keV$ suggest a magnetic field strength of $B\sim10^{5}\, \rm G$. However, all the best fit models underpredict the flux observed in the NIR wavelengths, which also only rebrightens by a factor of $\sim2$ during the second prompt emission episode, in stark contrast to the X-ray emission, which rebrightens by a factor of $\sim100$, suggesting an afterglow component is dominating the emission. We present GRB 121217A one of the few GRBs for which there are multi-wavelength observations of the prompt-emission period and show that it can be understood with a synchrotron radiation model. However, due to the complexity of the GRB's emission, other mechanisms that result in Band-like spectra cannot be ruled out.}


   \maketitle


\section{Introduction}
\label{sec:Introduction}

Ever since $\gamma$-ray bursts (GRBs) were first detected in the 1960s~\citep[][]{Klebesadel73a}, satellites have been launched to expand our understanding of the underlying mechanism that causes them. The most notable are the instrument BATSE~\citep[][]{Fishman89a}, onboard the {\it CGRO} satellite, and the {\it Swift}~\citep{Gehrels04a} and {\it Fermi}~\citep{Atwood09a} satellites, which were launched in the periods of 1990-2008. They have collectively detected thousands of long-duration GRBs and acquired many prompt-emission light curves and $\gamma$-ray spectra. Even though this huge data set has answered many questions about the GRB phenomenon, the underlying problem of the prompt-emission mechanism remains elusive~\citep[for a review, see][]{Zhang11c,Zhang12a}.

The standard model of a long-duration GRB involves a compact object, formed by the collapse of a massive rapidly rotating star~\citep{Woosley93a,Paczynski98a,MacFadyen99a}, that emits jetted relativistic fireballs with different Lorentz factors~\citep{Eichler89a,Narayan92a,Meszaros02a}. The most commonly discussed model of the prompt emission is the internal-shock scenario~\citep[e.g.,][]{Rees94a}, whereby the emitted fireball shells of different Lorentz factors cross one another causing relativistic shocks. Fermi acceleration~\citep{Fermi49a} across the shock front in combination with amplified magnetic fields results in the electrons cooling in the form of synchrotron radiation~\citep{Sari98a}, primarily at X-ray wavelengths, which is blueshifted into $\gamma$-rays~\citep[for a review, see, e.g.,][]{Meszaros02a, Zhang11a}. Such a scenario allows easy comparison with observations by fitting power laws to the observed spectra. However, the internal shock model predicts a relatively low radiative efficiency~\citep{Kumar99a, Panaitescu99a} and a wrong peak energy \citep[unless a small fraction of electrons are accelerated,][]{Daigne98a}. Recent numerical simulations suggest that internal shocks cannot efficiently accelerate particles if the ejecta carries a magnetic field~\citep[even if with a moderate magnetisation,][]{Sironi09a}. Also, a fireball giving rise to a strong internal shock emission is expected to have a bright quasi-thermal photosphere component, which is not observed as expected in some GRBs~\citep{Zhang09a}. As a result, alternative models of GRB prompt emission are widely discussed in the literature. These include 
magnetic dissipation models in a Poynting-flux-dominated flow~\citep[e.g.,][]{Zhang11a} or a dissipative photosphere model~\citep[e.g.,][]{Rees05a, Vurm11a}. 
To determine and constrain the preferred mechanism it is crucial to obtain multi-wavelength measurements during prompt-emission episodes of the GRB. 

Multi-wavelength measurements of the prompt-emission are not always possible, given the delay between the triggering of $\gamma$-ray telescopes and the slewing of optical instruments. Fortunately, however, there exist tens of fortuitous cases in which both the $\gamma$-ray emission and optical emission have been detected during the prompt period. These can be divided into three possible scenarios: (i) a wide-field camera is observing the same field position as a satellite and so catches the optical emission simultaneously~\citep[e.g., 080319B, 130427A;][]{Racusin08a,Bloom09a,Beskin10a,Wren13a}, (ii) the prompt period is long enough that optical instruments slew in time to observe the prompt period~\citep[e.g., 990123, 080928, 110205A, 091024;][]{Akerlof99a, Rossi11a, Cucchiara11a, Gruber11b, Gendre12a, Zheng12a, Virgili13a}, and (iii) there is a precursor to the main event so that optical instruments can slew in time~\citep[e.g., 041219A, 050820A, 061121;][]{Blake05a, Vestrand05a, Vestrand06a, PageK07a}. Only recently has it become possible to compile samples of bursts that exhibit optical emission during the prompt phase~\citep{Kopac13a}. However, their heterogeneous selection means that many more robust detections are required to reach both large number statistics and significant completeness levels.

Despite the successful efforts to detect the optical emission during the prompt episode, there is still no consistent picture on the underlying mechanism~\citep[e.g.,][]{Kopac13a}. Sometimes, optical emission of GRBs traces the $\gamma$-ray emission, but the optical emission is orders of magnitudes larger than expected by theory (e.g., 110205A), some do not trace the $\gamma$-ray emission (e.g., 990123), and some are below what is expected (e.g., 061121). A major problem is that, in the majority of cases, the observations are limited to only one filter or are not simultaneous observations, which makes it difficult to disentangle temporal and spectral variations. This highlights the importance of simultaneous multi-wavelength observations of the GRB prompt emission to investigate the underlying mechanism.

We add the {\it Swift}/{\it Fermi} burst GRB 121217A to this handful of cases, firstly by discussing its detections in Sect.~\ref{sec:Observations}. Secondly, we present the resulting light curves and spectra in Sect.~\ref{sec:Results}, discuss the implications in Sect.~\ref{sec:Discussion}, and then finally conclude in Sect.~\ref{sec:Conclusion}. Throughout we assume the standard notation of the GRB light curves and spectra of $F\left(\nu,t\right)\propto t^{-\alpha}\nu^{-\beta}$. In addition, we adopt the notation of $Q_{x}=10^{x}Q$. Unless mentioned otherwise, all uncertainties are quoted to the $1\sigma$ level. Finally, a $\Lambda$CDM cosmology with the following parameters $\Omega_{\Lambda}=0.7$, $\Omega_{\mathrm{M}}=0.3$, and $H_{0}=73.0\,\mathrm{km\,s^{-1}\,Mpc^{-1}}$ has been used~\citep{Freedman10a}. 

\section{Observations}
\label{sec:Observations}

\subsection{Swift}
\label{sec:Observations:subsec:Swift}
\begin{figure}
  \centering
  \includegraphics[clip=True,trim=1cm 0 0 0,width=9.5cm]{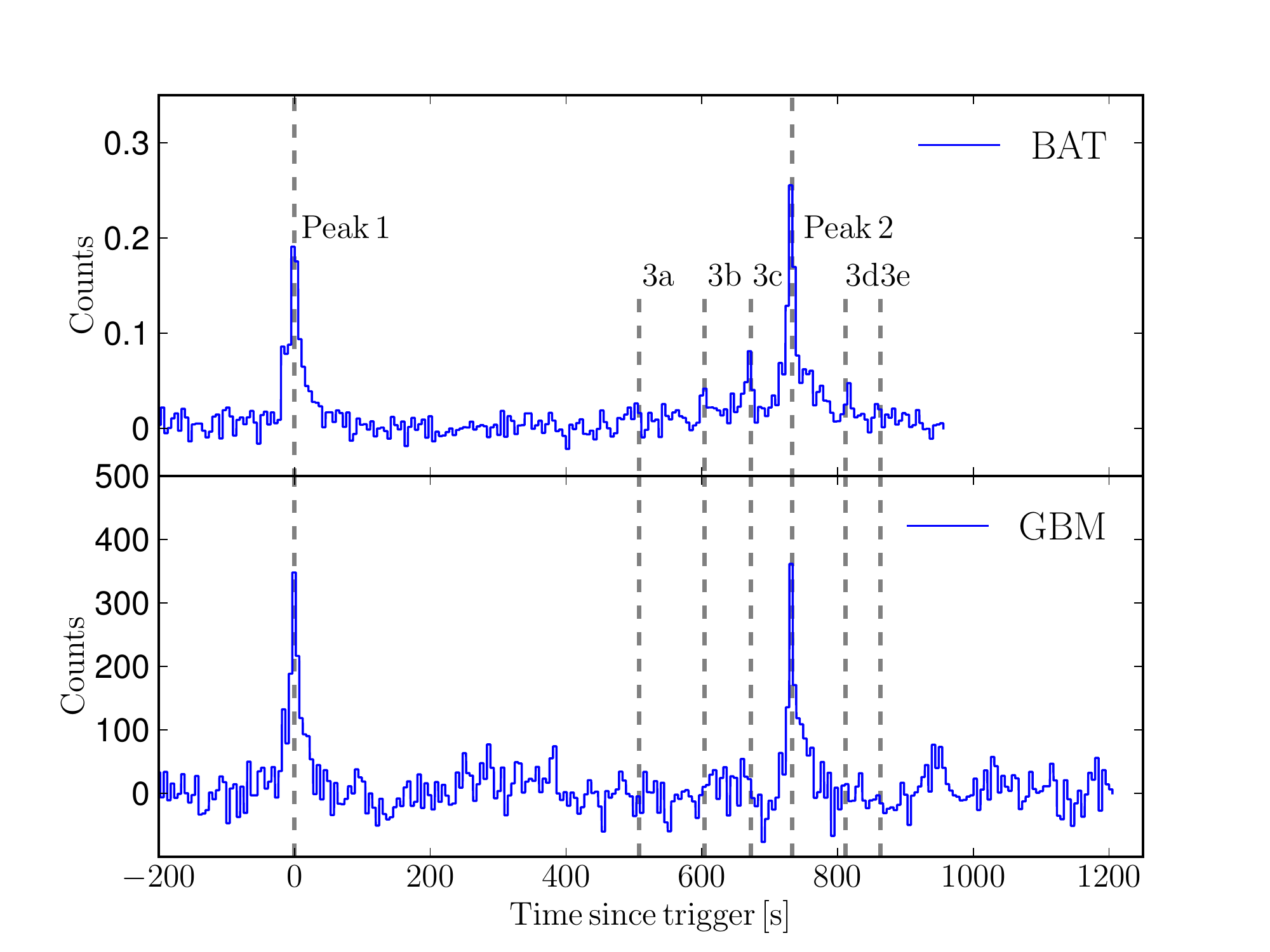}
  \caption{The $\gamma$-ray light curves of the two prompt episodes of GRB 121217A (Peak 1 and 2) acquired with BAT ($15-150\, \rm keV$) and GBM ($8-1000\, \rm keV$). The GBM triggered on the second peak which occurs at a time of $T_{0}+735\, s$, but has been shifted in this plot to coincide with the BAT $T_{0}$. Both light curves have been binned in time with a moving box of 5 seconds. There is extra emission seen before and after the second peak (3a, 3b, 3c, 3d, and 3e). The corresponding times and instrument detections are noted in Table~\ref{tab:peak_times}.}
  \label{fig:prompt_bat_and_gbm}
\end{figure}

\begin{figure*}
  \centering
  \includegraphics[clip=True,trim= 0 0 0 0, width=16cm]{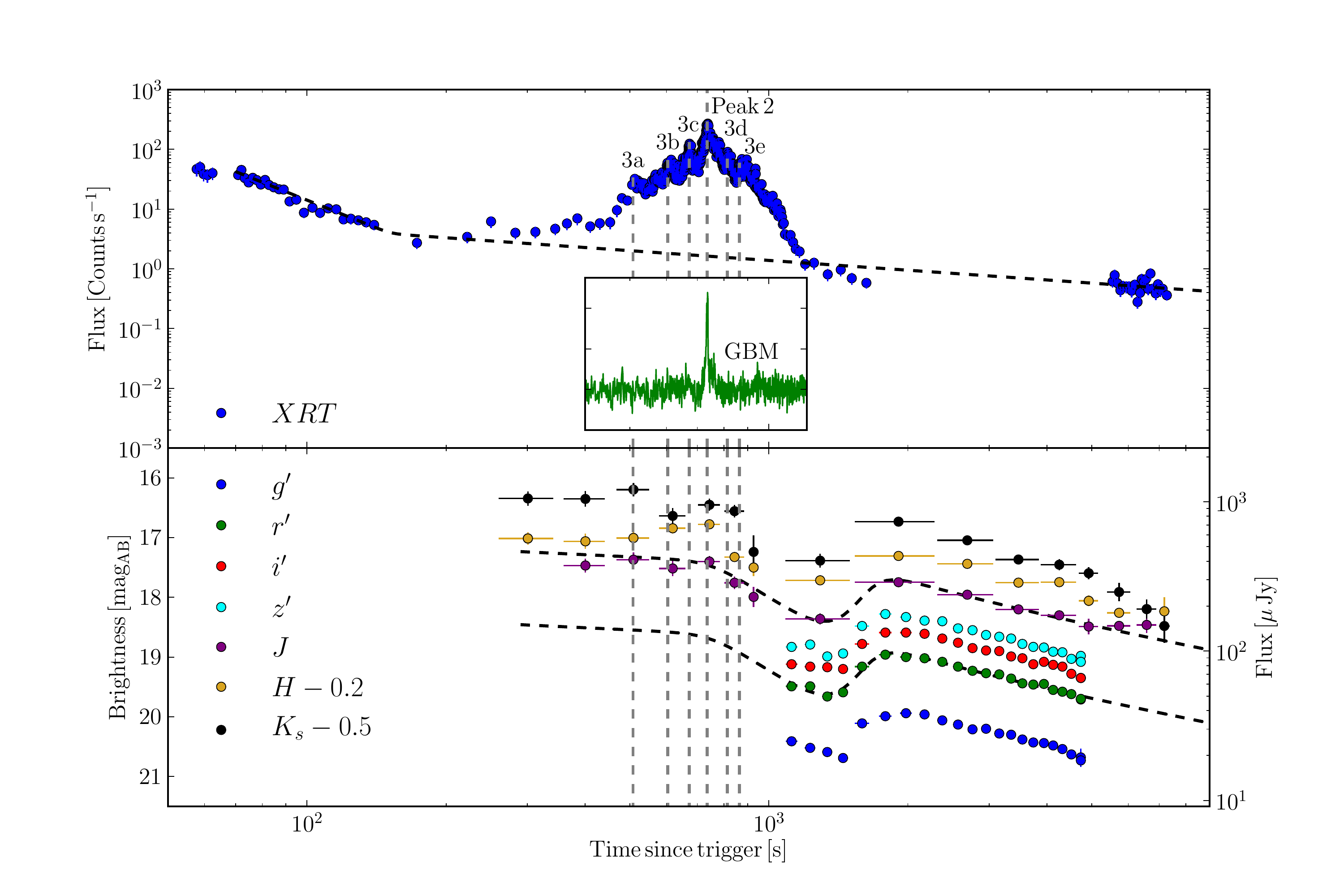}
  \caption{The X-ray (top) and optical/NIR (bottom) light curves of GRB 121217A, with the inset graph showing the GBM prompt emission of the second peak. The canonical model of the X-ray emission can be seen as the black-dashed line in the top panel, where the flare, simultaneous to the second prompt peak, has been excluded from the fit (Sect.~\ref{sec:Results:subsec:XrayEmission}). The black-dashed line in the lower panel is the best-fit double broken power law of the afterglow emission (Sect.~\ref{sec:Results:subsec:OpticalNIREmission}). Only the observations with a time less than $T_{0}+10^{4}\rm\, s$ have been included and the full light curve can be seen in Fig.~\ref{fig:lc_full}}.
  \label{fig:xrt_optical_lc_fits}
\end{figure*}

The Burst Alert Telescope ~\citep[BAT;][]{Barthelmy05a} mounted on {\it Swift} was triggered by GRB 121217A on 17th December 2012 at $T_{0}=$ 07:17:47 UT~\citep{Siegel12a, Cummings12a}. {\it Swift} slewed immediately to the burst and the X-Ray Telescope~\citep[XRT;][]{Burrows05a} began observing at $T_{0}+64.0\,\rm s$ until 15.6 days later~\citep{Evans12a}. The BAT light curve was acquired from the {\it Swift} quick-look data and the BAT spectrum was acquired from the {\it Swift} archive. The prompt light curve can be seen in Fig.~\ref{fig:prompt_bat_and_gbm}. The HEAsoft routines \emph{batbinevt}, \emph{bathotpix}, \emph{batmaskwtevt}, \emph{batupdatephakw}, and \emph{batdrmgen} were used to generate the BAT PHA and RSP files from the event file in the standard manner. The XRT light curve (Fig.~\ref{fig:xrt_optical_lc_fits}) was obtained from the XRT light-curve repository~\citep{Evans07a,Evans09a} and the XRT spectral data from the public {\it Swift} archive. Each spectrum has been regrouped to ensure at least 20 counts per bin using the \emph{grppha} task from the HEAsoft package using the response matrices from the CALDB \texttt{v20120209}. We assume a Galactic hydrogen column density of $0.4\times10^{22}\, \rm cm^{-2}$~\citep{Kalberla05a}.

The prompt emission exhibits two main emission periods separated by a quiescent period of $\sim500\,\rm s$ and lasts for a length of $\sim900\, \rm s$.
There are two main peaks which will be referred to as Peak 1 (first peak) and Peak 2 (second peak). There are also three smaller emission peaks prior to the second peak and two after, as depicted in Fig.~\ref{fig:prompt_bat_and_gbm}. Their corresponding names and times can be seen in Table~\ref{tab:peak_times}.

\begin{table}
  \begin{center}
    \caption{Times of $\gamma$-ray emission.}
      \begin{tabular}{c c c c c}
        \hline
        Name & Time$^{a}$ & Duration$^{b}$ & BAT$^{c}$ & GBM$^{c}$ \\
        \hline
             & s & s\\
        \hline
        Peak $1$ & $0$   & $69$ & Y & Y \\
        Peak $2$ & $735$ & $70$ & Y & Y \\
        $3\rm a$ & $508$ & $26$ & Y & N \\
        $3\rm b$ & $604$ & $19$ & Y & Y$^{d}$ \\
        $3\rm c$ & $635$ & $18$ & Y & N \\
        $3\rm d$ & $813$ & $36$ & Y & N \\
        $3\rm e$ & $863$ & $27$ & Y & N \\
        \hline
        \hline
      \end{tabular}
     \tablefoot{\tablefoottext{a}{All times are in reference to $T_{0}$ and are taken from BAT.}
          \tablefoottext{b}{All durations are obtained from BAT and refer to their length, not the $T_{90}$.}
     			\tablefoottext{c}{Detected in this instrument.}
     			\tablefoottext{d}{Excluded due to poor signal-to-noise.}}
  \label{tab:peak_times}
  \end{center}
\end{table}

\subsection{GBM}
\label{sec:Observations:subsec:GBM}
The {\it Fermi} Gamma-ray Burst Monitor~\citep[GBM;][]{Atwood09a} was triggered by Peak 2 on 17th December 2012 at 07:30:02 UT~\citep{Yu12a}. Even though GBM's triggering was switched off during the first peak of the prompt emission, as it was moving through a region of high geomagnetic activity, the first peak was still detected. The overall light curve resembles that seen with BAT (Fig.~\ref{fig:prompt_bat_and_gbm}). The GBM spectra were reduced in the standard manner using the \emph{RMFIT} \texttt{v4.1BA} software package and the Response Generator \emph{gbmrsp} \texttt{v2.0}. Fluences were determined using CSPEC data (time resolution of 4.096 s) and spectral fitting utilised time-tagged event data (time resolution of 64 ms).

\subsection{GROND}
\label{sec:Observations:subsec:GROND}
The Gamma-Ray burst Optical Near-infrared Detector~\citep[GROND;][]{Greiner08a} at the MPG/ESO 2.2 m telescope at La Silla, Chile, began observing the field of GRB 121217A at $T=T_{0}+210\,\rm s$ and located the optical/near-infrared (NIR) counterpart of GRB 121217A at R.A.(J2000) = $10^{h}14^{m}50.4^{s}$, Dec.(J2000) = $-62\degr 21\arcmin 0\farcs4$~\citep{Elliott12b} to an uncertainty of $0\farcs5$. As a result of the XRT position being at the edge of the BAT error circle, the afterglow was outside the field-of-view of the optical detectors for the first 500 s of observations. Therefore, there is only coverage in the NIR filters, due to the larger field-of-view of the NIR detectors. The telescope was then repointed, yielding a short interruption in the observations around 1000 s. The follow-up campaign lasted for 21 days until the afterglow was no longer detected. No underlying candidate host galaxy was discovered to a limit of $r'_{\rm AB}>24.9\, \rm mag$.

Image reduction and photometry of the GROND observations were carried out using standard IRAF tasks~\citep{Tody93a} in the way outlined in~\citet{Kruehler08a} and~\citet{Yoldas08a}. In brief, a point-spread function (PSF) was obtained from the bright stars within the field and applied to the afterglow. The absolute calibration of the optical photometry was achieved by observing a Sloan Digital Sky Survey (SDSS) field~\citep{Aihara11a} at R.A. (J2000) = $10^{h}50^{m}36.0^{s}$, Dec. (J2000) = $-21\degr 36\arcmin 00\arcsec$ and the GRB field consecutively, under photometric conditions. The NIR absolute calibration was obtained from the Two Micron Sky Survey (2MASS) stars~\citep{Skrutskie06a} within the field of the GRB. The magnitudes are corrected for a Galactic dust reddening of $E_{B-V}^{\rm Gal}=0.324\,\rm mag$ corresponding to an extinction of $A^{\rm Gal}_{V}=1.0\,\rm mag$ for $R_{V}=3.1$~\citep{Schlafly11a}. The magnitudes of GRB 121217A and its reference stars can be found in Tables~\ref{tab:lc_refstars} to \ref{tab:lc_nir}.

\section{Results}
\label{sec:Results}

\subsection{Redshift}
\label{sec:Results:subsec:Redshift}

\begin{figure}
  \centering
  \includegraphics[width=9.5cm]{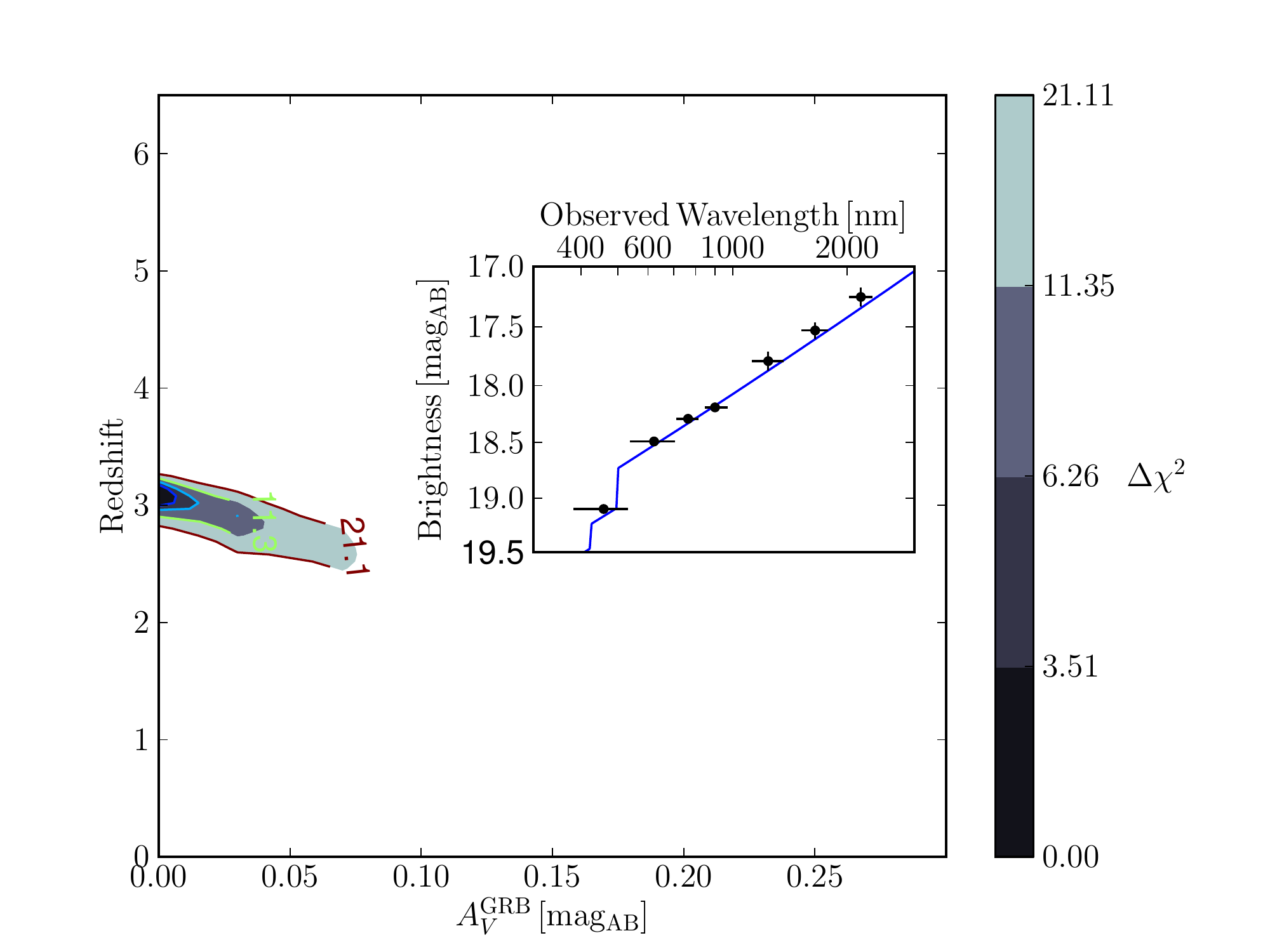}

  \caption{Contour plot of $\Delta\chi^{2}$ values for each of the host-galaxy dust ($A_{V}$) and GRB redshift ($z$) parameters used in the fitting grid (Sect.~\protect\ref{sec:Results:subsec:Redshift}), in comparison to the best-fit power law ($\beta=0.9$), and Small Magellanic Cloud dust model. The other dust model contour plots are not included, as they do not portray any different information. The inset is the broadband SED, the black dots are GROND data and the blue line depicts the best-fit power law. The significance levels for $\rm d.o.f.=3$ are: $1\sigma=68.3\%$, $\Delta\chi^{2}=3.5$, $2\sigma=90\%$, $\Delta\chi^{2}=6.3$, $3\sigma=99\%$, $\Delta\chi^{2}=11.4$, $4\sigma=99.99\%$, and $\Delta\chi^{2}=21.1$.
  }
  \label{fig:redshift_contour}
\end{figure}

A spectral energy distribution was constructed from the GROND filters at a mid-time of $T_{0}+31.4\,\rm min$, at which stage the optical counterpart is the afterglow component (see Sect.~\ref{sec:Results:subsec:OpticalNIREmission} for more details). Within the framework of the standard fireball model, external shocks emit synchrotron radiation, which is described by a (broken) power law~\citep[e.g.,][]{Sari98a}. These power laws are then modified by the GRB host's intrinsic extinction and the GRBs redshift, which determines the position of the Lyman-break. To find the redshift and intrinsic host dust extinction we followed the prescription outlined in~\citet{Kruehler10a} and fit power laws over a grid of parameters consisting of: spectral slope $\beta=0.01-2.00$ in steps of $0.01$, host dust $A_{V}=0.0-0.5\,\rm mag$ in steps of $0.02$, dust models (Milky Way, Large Magellanic Cloud and Small Magellanic Cloud), and redshift $z=0.0-5.0$ in steps of $0.06$. The best fit solution is determined from the minimum $\chi^{2}$ value and the uncertainties from the corresponding $\chi^{2}$ contours. We find a best fit solution with $\chi^{2}/\rm d.o.f.=3.7/3$ for the Small Magellanic Cloud dust model, $\beta=0.87^{+0.04}_{-0.07}$, $A_{V}=0.00^{+0.03}_{-0.00}\,\rm mag$, and $z=3.08^{+0.11}_{-0.06}$ to at least the $3\sigma$ level (Fig.~\ref{fig:redshift_contour}). The best fits for the other dust models return consistent results and we observe no change larger than $3\sigma$ in the host dust requirement if $30\%$ more or less of Galactic dust reddening is used. From here on any fits requiring redshift will be set to the best-fit value for simplicity and the intrinsic host galaxy dust absorption to zero. As there is no strong $2175\AA$ feature in the SED, we adopt the Small Magellanic Cloud best-fit, which is favoured for the majority of GRB afterglows~\citep[e.g.,][]{Kann10a}. The Large Magellanic and Milky Way dust models prefer host galaxy dust quantities of $A_{V}\sim0.1\, \rm mag$. However, they remain consistent with the Small Magellanic Cloud best-fit dust values at the $3\sigma$ level and do not change the results in the rest of the paper. 

No spectroscopic redshift has been reported. Ultraviolet detections would help to improve the redshift estimate. However, the upper limits during the afterglow phase from the Ultra-Violet Optical Telescope~\citep[UVOT;][]{Roming05a} onboard {\it Swift} are not constraining. Also, the detections of the afterglow yielded by co-adding over time frames of $\sim500\, \rm s$~\citep{Oates12a} have poor signal-to-noise and fold in the complexity of the light curve. As a result, the UVOT data do not provide tighter constraints on the redshift. At $z\sim3$, the Ly-limit is within the $u$-band and given the large errors, the $u$-band detection and UV-limits are thus consistent with the redshift derived from the GROND data.

\subsection{Host galaxy hydrogen column density}
\label{sec:Results:subsec:hydrogen}

The X-ray light curve behaves like a typical afterglow component at times of $t>T_{0}+1100\, \rm s$, showing no spectral evolution with a hardness ratio of $1.5\pm0.4$. We fit the X-ray data in the time range of $T_{0}+5496\, \rm s$ to $T_{0}+1.4\times10^{6}\, \rm s$ with a power law model and a fixed redshift of $z=3.08$, resulting in a best-fit hydrogen column density of $N_{\rm H,X}=2.1\pm0.8\times10^{22}\,\rm cm^{-2}$ with $\chi^{2}/\rm d.o.f.=116/111$. For consistency we adopt this value throughout the paper. We note the uncertainties are consistent with zero at the $3\sigma$ level. However, fixing $N_{\rm H,X}=0\, \rm cm^{-2}$ does not change the overall results outlined in this paper. 

\subsection{X-ray emission}
\label{sec:Results:subsec:XrayEmission}

We obtain the best fit X-ray light curve from the {\it Swift} online catalogue~\citep{Evans07a,Evans09a}, which determines the temporal slopes based on the type of classification that best fits the data, be it: canonical, one-break, no-breaks or undefined. By ignoring the flaring activity~\citep{Willingale07a} in the X-ray emission that occurs between $T_{0}+200\,\rm s$ and $T_{0}+5715\,\rm s$, the canonical afterglow reproduces the X-ray light curve the best (see Fig.~\ref{fig:xrt_optical_lc_fits}) with $\chi^{2}=166/146$~\citep[a canonical afterglow usually comprises of three power law segments, a fast initial decay of $3<\alpha<5$, followed by a shallow decay $0.5<\alpha<1.0$, and finishes with a slightly steeper decay $1<\alpha<1.5$,][]{Nousek06a}.

Starting at $T_{0}+72\,\rm s$, the X-ray light curve begins with a decaying power law with a pre-flare temporal slope of $\alpha_{1,\rm X}=3.14\pm0.18$ and a spectral slope of $\beta_{1,\rm X}=1.11\pm0.01$, between $T_{0}+72\rm s$ and $T_{0}+152\rm s$. The steep decay is then followed by an increase in X-ray emission, where the peak flux is simultaneous to the second prompt-emission peak. Both the flaring and second peak are discussed more thoroughly in Sect.~\ref{sec:Discussion:subsec:SecondPromptEmission}. After the flaring activity, the X-ray returns to a standard decay with a post-flare temporal slope of $\alpha_{2,\rm X}=0.54^{+0.05}_{-0.17}$ and spectral slope of $\beta_{2,\rm X}=0.92\pm0.06$. Directly after the X-ray peak the X-ray emission is systematically below the best-fit line. If achromatic, the X-ray light curve could have the same behaviour as the optical light curve, which is also decreasing at this time. However, there is no coverage of the X-ray emission during the optical rebrightening to place any constraint on the shape of the light curve and so it is possible that $\alpha_{2,\rm X}$ is underestimated. This decay is then followed by a break at $t_{\rm 3,b}=T_{0}+2.6\times10^{4}\,\rm s$, which steepens the decay to a final temporal slope of $\alpha_{3,\rm X}=1.38^{+0.06}_{-0.09}$ and spectral slope $\beta_{3,\rm X}=0.96\pm0.06$.

\subsection{Optical/NIR emission}
\label{sec:Results:subsec:OpticalNIREmission}

The NIR emission depicts no visible synchronous rebrightening during the second prompt emission and we defer the reader to later discussions (Sect.~\ref{sec:Discussion:subsec:SecondPromptEmission}). We assume that a second component, most likely an afterglow related to the first prompt emission, is dominating the NIR wavelengths. In the external shock model~\citep{Sari98a}, the afterglow phase is believed to be when the fireball begins to decelerate as it ploughs into the interstellar medium or progenitor winds, which results in a power-law decay~\citep[alternative models also require a power-law decay, e.g., the canon ball model;][]{Dar04a}. Therefore, we fit a double broken power law, of the~\citet{Beuermann99a} type, to the seven bands of GROND simultaneously. The best-fit solution seen in Fig.~\ref{fig:xrt_optical_lc_fits}, has $\chi^{2}/\rm d.o.f.=166/115$, which is high primarily due to the early and late NIR data, removing it yields a reduced-$\chi^{2}\sim1$. This suggests that the NIR uncertainties are being underestimated, but also could be a result of intrinsic variability as a result of flaring activity (see Fig.~\ref{fig:lc_zoom}). The NIR (and later optical) emission begins with a shallow decay with a temporal slope of $\alpha_{\rm opt,1}=0.15\pm0.03$ and then breaks at a time $T_{0}+(750\pm19)\,\rm s$ to a temporal slope of $\alpha_{\rm opt,2}=2.0\pm0.1$ with a smoothness $s_{\rm opt,12}=8.0\pm1.5$, a steep slope but expected from a reverse shock~\citep{Kobayashi03a}.
At a time of $\sim T_{0}+1450\,\rm s$ the optical emission begins to rebrighten with $\alpha_{\rm opt,3}=-1.8\pm0.2$ until it reaches a maximum at $T_{0}+(1669\pm10)\,\rm s$, with a smoothness of $s_{\rm opt,34}=9.8\pm0.2$, and once again begins to decay with $\alpha_{\rm opt,4}=0.59\pm0.02$. The spectral slope remains constant throughout with $\beta_{\rm opt,34}=0.90\pm0.04$. GROND coverage does not begin again until $T_{0}+8\times10^{4}\, \rm s$ but is consistent with a steeper slope of $\alpha_{\rm opt,5}=1.19\pm0.10$, consistent with the X-ray light curve.

\subsection{Broadband prompt-emission spectrum}
\label{sec:Results:subsec:PromptEmissionBroadBandSpectrum}

\begin{figure*}
  \begin{center}
    \includegraphics[width=16cm]{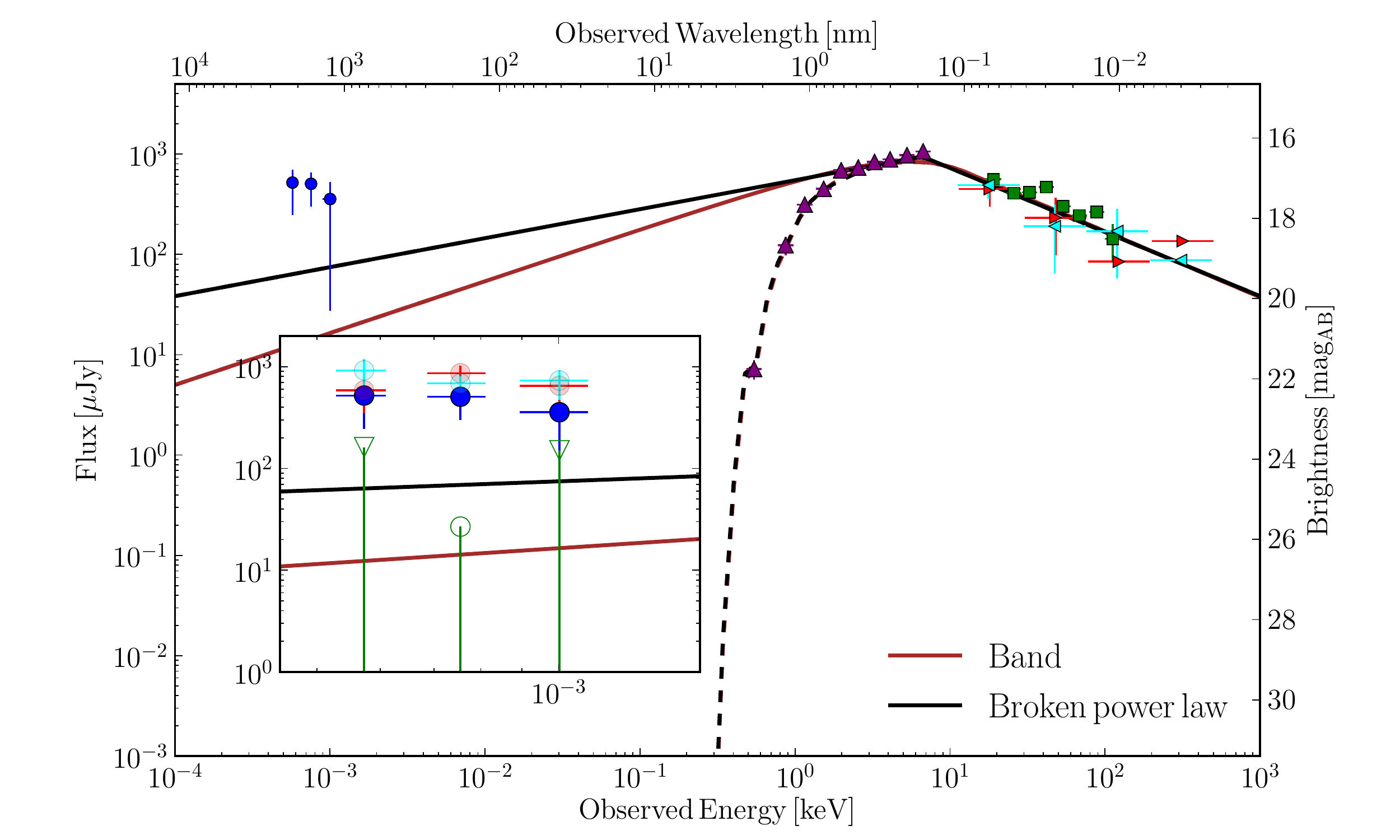}
    \caption{X-/$\gamma$-ray SED at the second prompt emission ($T_{0}+735\,\rm s$), composed of BAT (green squares), XRT (purple upward-triangles), and GBM's Na and N9 detectors (red rightward-triangles and cyan leftward-triangles, respectively). To make the plot more clear, the data points have been rebinned in energy and the BGO detectors non-detections are not included, but are consistent with the models. The best-fit models for the broken power law and Band models are depicted as the black and brown lines, respectively. The dashed lines correspond to the best-fit model with the effect of Galactic gas absorption. The inset shows a zoom in of the NIR wavelengths. The open green circle and triangles are with the expected afterglow flux subtracted, which was determined from the best-fit temporal power law. The transparent cyan and red dots denote the fluxes measured at $T_{0}+759$ and $T_{0}+769$ seconds, respectively, which can also be seen in Fig.~\ref{fig:lc_zoom}.}
    \label{fig:prompt_sed}
  \end{center}
\end{figure*}

We construct a broadband SED at the time of the second prompt emission (Peak 2), occurring at a mid-time of $T_{0}+(735\pm10)\,\rm s$, utilising the three NIR filters of GROND ($JHK$), {\it Swift}/BAT, {\it Swift}/XRT and {\it Fermi}/GBM (Fig.~\ref{fig:prompt_sed}). As mentioned above, the source was unfortunately located outside the field-of-view of the optical bands of GROND at this time. A common time interval of 10 seconds has been used because this is the minimum integration time of the GROND/NIR images that were taken. The spectra from X-ray to $\gamma$-ray wavelengths were fit in XSPEC \texttt{v12.7.1} with a power law (PL), \emph{pow}, a broken power law (BPL), \emph{bknpow}, and the Band function, \emph{grbm}, (both including and ignoring the NIR data) and each of the resulting best-fit parameters can be found in Table~\ref{tab:prompt_sed_fit} or are depicted in Fig.~\ref{fig:prompt_sed}. For fits which do not include NIR data, we extrapolate each of the models to the NIR wavelengths ($\sim1\,\rm \mu m$), as seen in the inset of Fig.~\ref{fig:prompt_sed}. 

\begin{table}
  \begin{center}
  \caption{The best-fit parameters for each spectral model for Peak 2.}
    \begin{tabular}{l c c l c }
      \hline
      Model & $\chi^{2}/$d.o.f. & $f^{a}$ & Parameter & Value \\
      \hline
      PL & $438/342$ & $0.07^{+0.02}_{-0.11}$ & $\beta$ & $0.32\pm0.02$ \\
         &           &                        & $E_{\rm iso}^{c}$ [$10^{53}\, \rm erg$] & $3.3$ \\
     \hline
      PL$^{b}$ & $676/345$ & $1.0^{+0.3}_{-1.7}$ & $\beta$ & $0.14\pm0.08$ \\
               &           &                     & $E_{\rm iso}^{c}$ [$10^{53}\, \rm erg$] & $6.2$ \\

      \hline
      \hline
      BPL & $306/340$ & $10^{+4}_{-3}$ & $\beta_{1}$ & $-0.29\pm0.06$ \\
          &           &   & $\beta_{2}$ &  $0.64\pm0.05$ \\
          &           &   & $E_{\rm b}$ [keV]& $6.60\pm0.85$ \\
          &           &   & $E_{\rm iso}^{c}$ [$10^{53}\, \rm erg$] & $1.7$ \\

      \hline
      BPL$^{b}$ & $316/343$ & $4^{+2}_{-2}$ & $\beta_{1}$ & $-0.18\pm0.05$ \\
                &           &   & $\beta_{2}$ &  $0.63\pm0.06$ \\
                &           &   & $E_{\rm b}$ [keV]& $7.13\pm1.14$ \\
                &           &   & $E_{\rm iso}^{c}$ [$10^{53}\, \rm erg$] & $1.7$ \\

      \hline
      \hline
      Band & $310/340$ & $48^{+175}_{-12}$  & $\beta_{1}$ & $-0.48\pm0.13$ \\
           &           &   & $\beta_{2}$ & $-1.66\pm0.06$ \\
           &           &   & $E_{\rm b}$ [keV] & $11\pm188$  \\
           &           &   & $E_{\rm iso}^{c}$ [$10^{53}\, \rm erg$] & $0.9$ \\
 
      \hline
      Band$^{b}$ & $327/343$ & $23^{+114}_{-7}$ & $\beta_{1}$ & $-0.58\pm0.11$ \\
                 &           &   & $\beta_{2}$ & $-1.65\pm0.06$ \\
                 &           &   & $E_{\rm b}$ [keV] & $14\pm8$  \\
                 &           &   & $E_{\rm iso}^{c}$ [$10^{53}\, \rm erg$] & $0.7$ \\

      \hline
      \hline
      \end{tabular}
     \tablefoot{\tablefoottext{a}{The ratio of the observed $J$-band flux to the expected $J$-band flux, i.e., $f=J_{\rm obs}/J_{\rm exp}$.}
                \tablefoottext{b}{Fits that have included the NIR channels.}
                All fits have assumed a redshift of $z=3.08$ and $N_{\rm H,X}=2.1\times10^{22}\, \rm cm^{-2}$ for simplicity, see Sects.~\ref{sec:Results:subsec:Redshift} and~\ref{sec:Results:subsec:hydrogen}.
		\tablefoottext{c}{Isotropic-equivalent energy calculated over the range of $0.1\, \rm keV$ to $10^{4}\, \rm keV$~\citep[see, e.g.,][]{Elliott12a}.}
	}

      \label{tab:prompt_sed_fit}
    \end{center}
\end{table}

\section{Discussion}
\label{sec:Discussion}

\subsection{High latitude emission: Deceleration radius and Lorentz factor of the first prompt peak}
\label{sec:Discussion:subsec:HighLatitudeEmission}

We consider the initial steep decay of the X-ray emission (Sect.~\ref{sec:Results:subsec:XrayEmission}), which is usually associated with high latitude emission~\citep[see, e.g.,][]{Zhang06a} and compare the fitted values to the expected closure relations~\citep{Kumar00a}. We find that $\alpha_{1,\rm closure,X}=2+\beta_{1,\rm X}=3.11\pm0.01$ (cf. $\alpha_{1,\rm X}=3.14\pm0.18$) and therefore this phase is consistent with being related to the prompt emission (i.e., high-latitude emission) and not the afterglow component. We also calculated the closure relations for an afterglow component~\citep[e.g.,][]{Sari98a,Racusin09a}, for $p>2$ within a Wind/ISM environment for fast/slow cooling and find that they cannot reproduce the temporal slope $\alpha_{1, \rm X}$ to at least $3\sigma$. Finally, the decay index is too steep to be a standard reverse~\citep[$\alpha\sim2$;][]{Kobayashi03a} or forward shock~\citep[$\alpha\sim1$;][]{Sari98a}.

Knowledge of the end time of the high-latitude emission allows us to estimate the radius at which the $\gamma$-rays originate and the Lorentz factor of the shell. We place a limit on the radius at which this emission occurs~\citep[e.g.][]{Lazzati06a,Meszaros06a,Zhang06a}, $R_{\gamma}$, with the following relation:

\begin{equation}
  t_{\rm tail} \lesssim (1+z)\frac{R_{\gamma}}{c}\frac{\theta_{\rm jet}^{2}}{2}.
\end{equation}

\noindent where $c$ is the speed of light, and $\theta_{\rm jet}$ is the jet half-opening angle. Using the time at which there is a canonical jet break in the X-ray emission, at $t_{3,\rm b}=T_{0}+2.6\times10^{4}\,\rm s$ (Sect.~\ref{sec:Results:subsec:XrayEmission}), the redshift $z=3.08$ (Sect.~\ref{sec:Results:subsec:Redshift}) and the isotropic-equivalent energy $E\sim10^{53}\,\rm erg$ (see Table~\ref{tab:prompt_sed_fit}) from the first peak, we can estimate the opening angle as $\theta_{\rm jet}=1.6\degr\,\left(\frac{n_{\gamma}}{0.2}\right)^{\frac{1}{8}}\left(\frac{n_{0}}{0.1\, \rm cm^{-3}}\right)^{\frac{1}{8}}$~\citep{Frail01a}. Assuming that the $\gamma$-ray efficiency, $n_{\gamma}=0.2$, and the ISM density, $n_{0}=1\,\rm cm^{-3}$, results in a prompt emission radius of

\begin{equation}
    R_{\gamma} \gtrsim \frac{2ct_{\rm tail}}{\left(1+z\right)\theta_{\rm jet}^{2}} = 1.6\times10^{15}\, \rm cm.
\end{equation}

\noindent This value could be smaller by a factor of a few if the break in the X-ray light curve is not the jet break. The emission from the high-latitude component is much brighter than the onset of the afterglow, which is not seen until the canonical plateau phase begins and so places an upper limit on the time ($t_{\rm dec}$), and thus the radius ($R_{\rm dec}$), at which deceleration of the shock begins. Utilising the fact that the tail emission ends at $T_{0}+152\,\rm s$ (i.e., $t_{\rm tail}=152\, \rm s$), we constrain the bulk Lorentz factor, $\Gamma_{0}$, by applying Eq.~6 of \citet{Zhang06a}:

\begin{equation}
  \Gamma_{0} \approx 1328 \left[\frac{E_{\gamma,\rm iso,52}\left(1+z\right)^{3}}{\left(\frac{n_{\gamma}}{0.2}\right)\left(\frac{n_{0}}{1\,cm^{-3}}\right)t_{\rm peak}^{3}}\right]^{\frac{1}{8}} \approx 487,
  \label{eqn:bulk_lf}
\end{equation}

\noindent where we have assumed $t_{\rm peak}=152\, \rm s$ and used the same constants as previously noted. We note that the tail emission could last much longer if the plateau phase is not a result of a rising afterglow component, which would reduce the estimate of the Lorentz factor. For example, using a time of $1000\, \rm s$ would halve the Lorentz factor.

After Peak 1 there are no spectral or temporal slopes that satisfy the high latitude emission closure relations to at least $3\sigma$ and this period is most likely masked by the complexity of the flaring activity.

\subsection{Optical afterglow rebrightening: Deceleration radius and Lorentz factor of the second prompt peak}
\label{sec:Discussion:subsec:OpticalAfterglow}

The optical afterglow-like component that is observed from $T_{0}+1670\, \rm s$ decays with a temporal slope of $0.59\pm0.02$ and a spectral slope of $0.87^{+0.04}_{-0.07}$. These combinations are not consistent with the standard closure relations for either an ISM or Wind environment for any of the frequency ranges to the $3\sigma$ level. Assuming an ISM environment in the slow cooling regime, with a frequency located at $\nu_{\rm m}<\nu<\nu_{\rm c}$, would require that $\alpha_{\rm closure}=\frac{3\beta_{\rm opt,34}}{2}=1.31^{+0.06}_{-0.11}$, much steeper than that observed. The shallow decay of the afterglow could be attributed to an injection of energy, consistent with the internal shock shells from the X-ray flaring activity that catch up with the primary forward shock. Assuming an injection of the form $E\propto t^{e}$~\citep{Panaitescu06a}, the difference in slopes of $\Delta \alpha=\alpha_{\rm closure}-\alpha_{\rm opt,\, 4}=1.31-0.59=0.72$ must satisfy the relation $\Delta \alpha=e\times1.36$ in an ISM environment~\citep{Panaitescu06a}. Therefore, the flatter slope can be explained with an injection parameter of $e=0.53$.

The rise time of the afterglow component, whether it be the forward or reverse shock, can be used to estimate the bulk Lorentz factor of the ejecta at the deceleration radius. We treat this shell as a \emph{thin shell}, regardless of whether it is associated to the first or second prompt emission, as in both cases there is a clear delay between the $\gamma$-ray emission and the maximum of the afterglow. Using equation~\ref{eqn:bulk_lf},

\begin{equation}
  \Gamma'_{0} \approx 247,
  \label{eqn:bulk_gamma}
\end{equation}

\noindent where any primed value is related to the second prompt emission. Therefore, the deceleration radius is

\begin{equation}
  {R'}_{\rm dec} = 2c{\Gamma'}_{0}^{2}t'_{\rm peak} = 3.4 \times 10^{18}\, \rm cm,
  \label{eqn:emission_radius}
\end{equation}

\noindent where we have assumed that the afterglow component is a result of the second peak, which occurs at $t'_{\rm peak}=934\, \rm s$. We have used the same fixed parameters as outlined in the previous section.

\subsection{NIR rebrightening during prompt emission}
\label{sec:Discussion:subsec:NoRebrightening}
\begin{figure}
  \begin{center}
    \includegraphics[width=9cm]{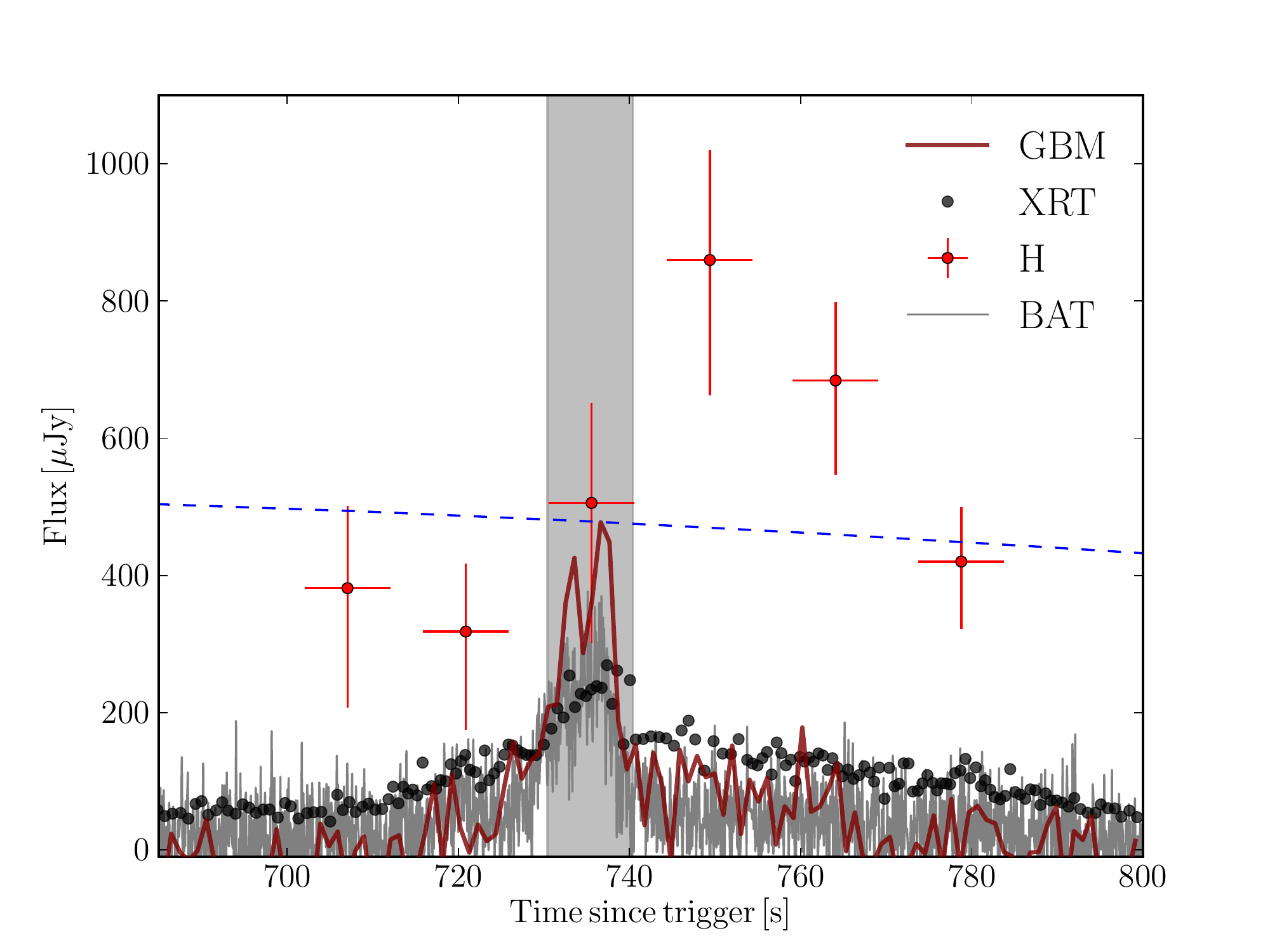}
    \caption{High time-resolution light curve of the second prompt emission of GRB 121217A, including BAT, GBM, XRT and the $H$-band of GROND. The best-fit $H$-band afterglow light curve from Sect.~\ref{sec:Results:subsec:OpticalNIREmission} is shown by the dashed blue line.
}
    \label{fig:lc_zoom}
  \end{center}
\end{figure}

The high time-resolution light curve seen in Fig.~\ref{fig:lc_zoom} shows that there is a rebrightening in the NIR wavelengths by a factor of $2.7\pm0.6$, but this is delayed from the peak of the X-ray/$\gamma$-ray emission by $(14\pm7)\, \rm s$. This change in flux is incredibly small when compared to the rebrightening in the X-rays, which changes by a factor of $\sim100$. 

A sample of eighteen bursts from~\citet{Kopac13a} that have optical/NIR coverage, in a single band, during the prompt phase, exhibit temporal slopes that are on average $\alpha<-5$. This is in stark contrast to the change in flux of GRB 121217A, which reinforces the idea that the flux from the prompt emission is being outshone by the afterglow emission.

The delay can be explained by considering two different internal-shock components or two shock emissions in a single internal shock component. In a simple internal shock model~\citep[the random shell model, e.g.,][]{Kobayashi97a}, an internal shock is described by a collision of two shells. Two shocks propagate in the outer and inner shell and if the two shells have very different mass densities the typical frequencies of the emission from the two shocks would be quite different. They would then peak at different times.

\subsection{Synchrotron radiation model}
\label{sec:Discussion:subsec:SecondPromptEmission}

The extension of the best-fit power law overpredicts the flux in the NIR wavelengths by a factor $\sim10$ (Table~\ref{tab:prompt_sed_fit}) and has a reduced-$\chi^{2}$ that is more than $3\sigma$ away from a perfect fit (reduced-$\chi^{2}=1$) and so we ignore it from here on. Both the broken power law and Band model fits have a reduced-$\chi^{2}$ that is within $3\sigma$ to a value of 1. Within synchrotron radiation theory the break in the power law could exist for two reasons: (i) fast/slow cooling of the electron population or (ii) photon self-absorption by the electron population.

\subsubsection{Fast and slow cooling}
\label{sec:Discussion:subsec:SecondPromptEmission:subsubsec:FastCooling}

The synchrotron radiation model predicts that electrons will have a spectral slope of $\nu^{\frac{1}{3}}$ below the maximum injection frequency $\nu_{m}$. The slope above this frequency is governed by the rate at which the electrons cool. If they cool faster than the dynamical time of the shock they will have a slope of $\nu^{-\frac{1}{2}}$ and if they are slower than the dynamical time then a slope of $\nu^{-\frac{(p-1)}{2}}$ where $p$ is the electron distribution power law index~\citep{Sari98a}.

The best-fit broken power laws (Table~\ref{tab:prompt_sed_fit}) below the best-fit break energy have slopes of $\beta_{1}=-0.29\pm0.06$ and $\beta_{1}=-0.18\pm0.05$ respectively and are consistent with a slope of $\nu^{\frac{1}{3}}$. Also, the slopes above the break frequency, $\beta_{2}=0.64\pm0.05$ and $\beta_{2}=0.63\pm0.06$, are still consistent with a slope of $\nu^{-\frac{1}{2}}$ to the $3\sigma$ level, assuming fast cooling. In addition, above the break frequency, a slope of $\beta=0.64$ would correspond to an electron index of $p=2.3$ for the slow cooling regime, which is a reasonable value in comparison to theory and observations~\citep[e.g.,][]{Rossi11a}.

Unfortunately, the broken power law under predicts the flux expected from the NIR emission and in combination with the small rebrightening in the NIR during the second prompt emission would suggest a secondary component is dominating the emission, most likely afterglow emission related to the first prompt-emission shell, as we have already noted~\citep[for another example, see, e.g.,][]{Kruehler09a}. We subtract the flux of the best-fit power law to the NIR light curve (see Sect.~\ref{sec:Results:subsec:OpticalNIREmission} or Fig.~\ref{fig:lc_zoom}), which we attribute to an afterglow component, during the second prompt-emission and plot the corresponding flux of the GROND $JHK$ bands in Fig.~\ref{fig:prompt_sed}. This places the observed spectrum in good agreement with the synchrotron radiation model (or any other model with a similar spectral index).

\subsubsection{Synchrotron self-absorption frequency}
\label{sec:Discussion:subsec:SecondPromptEmission:subsubsec:SelfAbsorption}

Inclusion of a secondary afterglow component is required to explain the excess flux of the NIR in combination with synchrotron radiation. However, the absorption frequency, $\nu_{a}'$, may also be between the optical and X-ray frequencies that is hidden underneath the secondary emission component. Self-absorption occurs when the optical wavelength photons are being absorbed by the radiating electrons and this occurs at the self-absorption frequency, $\nu'_{\rm a}$. We consider the two scenarios outlined in~\citet{Shen09a}, where the self-absorption frequency is below the optical frequency, ${\nu_{\rm a}'}<{\nu_{\rm opt}'}<{\nu_{\rm p}'}$~\citep[case III of][]{Shen09a}, or the self-absorption frequency is between the optical and X-ray observing frequencies, ${\nu_{\rm opt}'}<{\nu_{\rm a}'}<{\nu_{\rm X}'}$~\citep[case IV of][]{Shen09a}. These frequencies are set by properties of the initial fireball, mainly the emission radius, $R'_{\gamma}$, the Lorentz factor, $\Gamma'$, and the magnetic field, $B'$. There is no visible spectral break in the prompt emission to define ${\nu_{\rm p}}$ and so at this stage we can only set an upper limit of ${\nu_{\rm p}}>2.4\times10^{20}\, \rm Hz$, with an emission of ${f'}_{\rm {\nu_{p}'}}=40\, \rm \mu Jy$ assuming the best-fit power law. Using equation 8 from~\citet{Shen09a} for case III, this would place a constraint on the self-absorption frequency of

\begin{equation}
{\nu'}_{\rm a} = \left[\frac{{f'}_{{\nu_{\rm p}'}}}{{f'}_{{\nu_{\rm opt}'}}}\left(\frac{{\nu_{\rm p}'}}{{\nu_{\rm m}'}}\right)^{\beta}{\nu_{\rm opt}'}^{2}{\nu_{\rm m}'}^{-\frac{1}{3}}\right]^{\frac{3}{5}} = 2.6\times10^{14}\, \rm Hz,
\end{equation}

\noindent where the measured flux in the $H$-band without the afterglow component is ${f'}_{\nu_{\rm opt}}=27\, \rm\mu Jy$,  ${\nu_{\rm opt}'}=1.8\times10^{14}\, \rm Hz$, ${\nu_{\rm m}'}=1.6\times10^{18}\, \rm Hz$, and $\beta=0.64$ from the best fit power law. We have also assumed the self-absorption is optically thin (i.e., ${\nu_{\rm opt}'}^{2}$), but the optically thick case is equally as valid (i.e., ${\nu_{\rm opt}'}^{\frac{5}{2}}$). The determined self-absorption frequency satisfies the relation ${\nu_{\rm opt}'}<{\nu_{\rm a}'}<{\nu_{\rm X}'}$ for the optically thin absorption, but not the optically thick. In the optically thin case the emission radius can be constrained, using~\citet{Shen09a} equations 12 and A17, to

\begin{equation}
  {R_{\gamma}'} = 4.3 \times 10^{14} {\Gamma'}_{300}^{\frac{3}{4}}{B'}_{5}^{\frac{1}{4}}\, \rm cm,
  \label{eqn:emission_radius_selfabsorption}
\end{equation}

\noindent using the same values outlined previously. This value is of the order of those determined in other works~\citep[e.g.,][]{Shen09a}. Substituting the Lorentz factor (Eq.~\ref{eqn:bulk_gamma}) into Eq.~\ref{eqn:emission_radius_selfabsorption} results in an estimate of the required magnetic field of

\begin{equation}
  {B'}_{5} \sim 1.2 {R'}^{4}_{\gamma,14}\, \rm G.
\end{equation}

\subsection{Flaring activity}
\label{sec:Discussion:subsec:BandModel}

\begin{figure}
  \begin{center}
    \includegraphics[width=9cm]{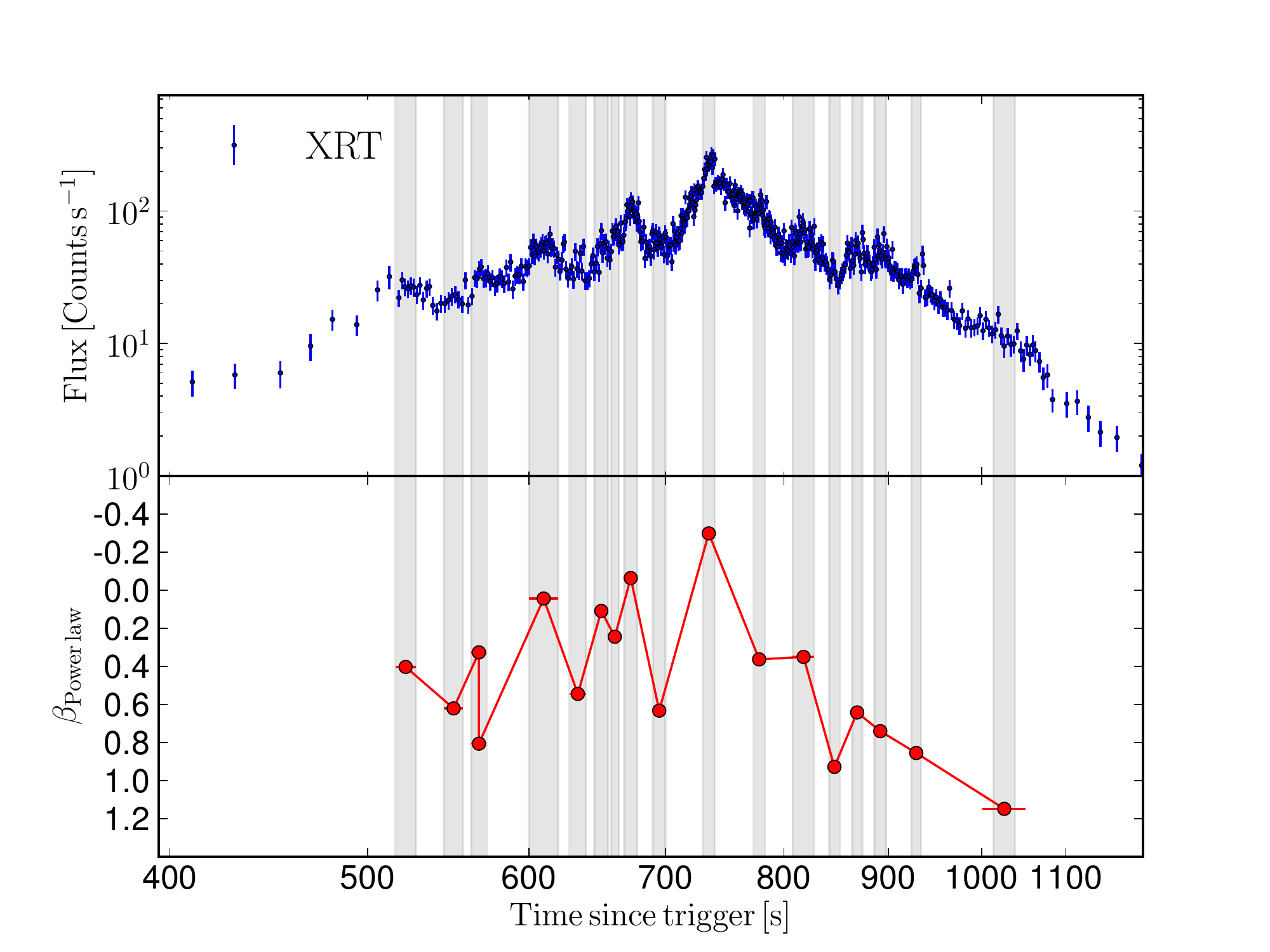}
    \caption{The top panel is a zoom in of the X-ray light curve during the second prompt emission that peaked at $t=T_{0}+735\, \rm s$. The bottom panel shows the spectral index of the best-fit power law, $\beta$, for time-sliced spectra shown by the grey shaded bars. The best-fit host gas absorption from the complete prompt SED has been assumed.}
    \label{fig:lc_xrt_ph}
  \end{center}
\end{figure}

The rebrightening period of the X-ray emission around the second $\gamma$-ray peak is shown in Fig.~\ref{fig:lc_xrt_ph}, during which several pulses and dips are evident. It has a maximum peak at the same time as the prompt emission. As already noted, for each of the main pulses in the X-ray emission there is an associated $\gamma$-ray pulse. The length of activity at X-ray wavelengths is $t\sim600\, \rm s$, whereas each of the prompt pulses lasts for $t\sim10-50\, \rm s$, at least twenty times shorter (see also Fig.~\ref{fig:xrt_optical_lc_fits}). We fit a power law for the times at which the X-ray light curve shows bumpy features to obtain the spectral slopes. 

The power law spectral index begins at $\beta\sim0.4$ and then approaches very flat values during each of the bumpy features and is the most flat at the time the prompt emission occurs, after which it becomes spectrally soft again, settling at $\beta\sim1.0$. Each of these bumps could be the result of slow shells (of low Lorentz Factors) causing internal shocks. 

\begin{figure}
  \begin{center}
    \includegraphics[width=9cm]{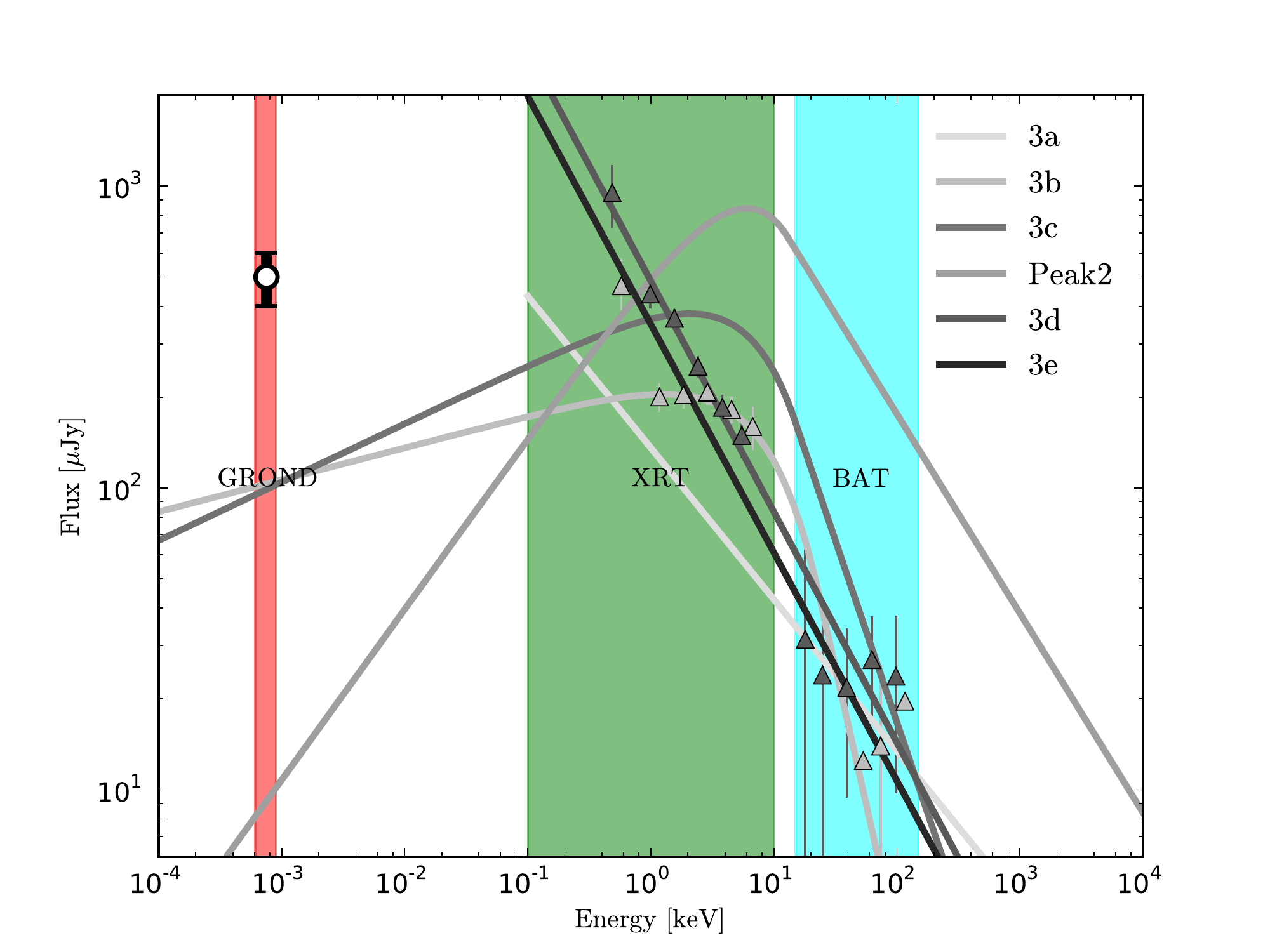}
    \caption{Band functions for the six prompt emission peaks. Each coloured column shows the wavelength coverage of the instruments (we neglect GBM for clarity as it covers parts of the XRT and BAT). They show that as the peak energy moves to lower energies the flux in the $\gamma$-ray wavelengths decreases substantially in comparison to the higher peak energy. It can be seen that a higher flux in $\gamma$-rays corresponds to a lower flux in the optical/NIR regime. No constraint on the peak energy was possible for 3a, 3d, and 3e, because they could lie anywhere below 0.1 keV and so were refit with a power law. The open circle depicts the (unsubtracted) $J$-band flux measured with GROND during Peak 2. The filled triangles are the X-ray and BAT $\gamma$-ray data and for reasons of clarity, we only show them for peaks 3b and 3d. The data has been corrected for the Galactic and host hydrogen absorption from the best-fit models. The best-fit parameters and goodness-of-fit for each of the peaks can be found in Table~\ref{tab:peak_fit}.}
    \label{fig:lc_xrt_bf}
  \end{center}
\end{figure}

In addition, we fit the other prompt emission peaks, 3a, 3b, 3c, 3d, and 3e with broken power laws and Band functions (see Fig.~\ref{fig:prompt_bat_and_gbm}) and plot them alongside the best-fit spectrum of the second prompt emission in Fig.~\ref{fig:lc_xrt_bf}. The spectra for which the frequency break or peak energy cannot be constrained are only plotted up to $0.1\, \rm keV$. As already noted, we can think of each bump as being an internal shock of fireballs with varying Lorentz factors. Each peak is then associated with a different maximum injection frequency $\nu_{m}$. The larger the $\nu_{\rm m}$ of the shock, the brighter it is in the $\gamma$-rays (neglecting any changes of spectral slope), with the brightest being detectable by the $\gamma$-ray telescopes. The lower $\nu_{m}$ is, the dimmer the $\gamma$-ray emission is (i.e., peaks 3a-e) and it is possibly not detected at all, while still remaining bright at X-ray wavelengths (i.e., peaks that seen in the X-ray emission but not in the $\gamma$-rays). For example, the brightest X-ray and $\gamma$-ray emission is Peak 2, which also has the largest injection frequency. However, we note that is also possible to explain all the features with a simple Band-like spectrum where the injection frequency described previously would be replaced by the peak energy ($E_{\rm peak}$) that is associated to every flare that occurs.

\begin{table*}
  \begin{center}
    \caption{The best-fit parameters for each peak.}
    \begin{tabular}{l c l l l}
    \hline
     Name & Model & $\chi^{2}$/d.o.f. & Parameter & Value \\
    \hline
      3a   & PL$^{a}$    & $91/88$ &  $\beta$ & $0.50\pm0.06$ \\
    \hline
      3b   & Band  & $58/77$ & $\beta_{1}$ & $-0.90\pm0.20$ \\
           &       &         & $\beta_{2}$ & $-2.82\pm1.10$ \\
           &       &         & $E_{\rm b}\, \rm [keV]$ & $11\pm33$ \\
    \hline
      3c   & Band  & $147/98$ & $\beta_{1}$ & $-0.81\pm0.14$ \\
           &       &          & $\beta_{2}$ & $-2.20\pm0.26$ \\
           &       &          & $E_{\rm b}\, \rm [keV]$ & $11\pm240$ \\
    \hline
      3d   & PL    & $81/85$ &$\beta$ & $0.76\pm0.06$ \\
    \hline
      3e   & PL    & $81/78$ &$\beta$ & $0.76\pm0.07$ \\
    \hline\hline 
    \end{tabular}

    \tablefoot{\tablefoottext{a}{When the Band model returned two slopes of the same value, the data were refit using a power-law model.}
              }
  \label{tab:peak_fit}
  \end{center}
\end{table*}

\section{Conclusions}
\label{sec:Conclusion}

The {\it Swift/Fermi} burst GRB 121217A was observed with two satellites and one ground-based telescope, with five different instruments covering the NIR, X-ray, and $\gamma$-ray wavelengths during a secondary prompt-emission period. The NIR emission exhibits a rebrightening during the prompt episode, that is much smaller in comparison to cases such as the Naked-Eye Burst (080319B). However, the X-ray emission increases by a factor of a hundred and exhibits several X-ray/$\gamma$-ray flares. 

The X-ray/$\gamma$-ray spectrum of the second prompt emission is well described by a broken power law, but underpredicts the flux expected in the NIR wavelengths. Attributing the additional emission to an afterglow component associated with the first prompt emission and subtracting its contribution would give a flux consistent with an extrapolation of the prompt spectrum of the GRB. In terms a synchrotron radiation model, the break frequency is interpreted as the maximum injection frequency, $\nu_{\rm m}$, at an energy of $E_{\rm m}=6.6\, \rm keV$. The possibility of the self-absorption frequency existing between the X-ray and optical frequencies allows the radius to be constrained to ${R_{\gamma}'} = 4.3 \times 10^{14} {\Gamma'}_{300}^{\frac{3}{4}}{B'}_{5}^{\frac{1}{4}}\, \rm cm$. Estimates of the bulk Lorentz factor obtained from the peak of the afterglow emission of $\Gamma_{0}\sim300$ and assuming a standard emission radius in an internal shock model of $R\sim10^{14}\, \rm cm$ would suggest a magnetic field strength of $B\sim10^{5}\, \rm G$. A photometric redshift of $z=3.1\pm0.1$ is determined from the afterglow emission.

Finally, the X-ray emission has several flaring episodes both prior and post to the prompt emission, showing that the central engine is active even after the initial prompt emission. The flaring can be explained in terms of the internal shock model as the collision of several fireball shells (with varying Lorentz factors) that are then easily described by synchrotron radiation theory. The lack of any large variability in the NIR wavelengths is then a result of a combination of the (i) dominating afterglow component, (ii) the synchrotron cooling slope of $\nu^{\frac{1}{3}}$, and (iii) possible self-absorption by the electron population. 

We presented GRB 121217A that was observed simultaneously in multiple NIR filters and by X-ray and $\gamma$-ray telescopes during its prompt emission and show that it can be explained in the framework of the internal shock model. Further observations that have high-time resolution ($\sim10\, \rm s$) with high signal-to-noise in the optical/NIR wavelengths, achievable at such facilities as the Very Large Telescope, in combination with other space-bound facilities that allow a wavelength coverage of $10^{-3}\, \rm keV$ to $10^{3}\, \rm keV$ are required to further distinguish between the underlying prompt-emission mechanism: internal shock model, magnetic dissipation, Poynting-flux dominated.

\begin{acknowledgements}
We thank the anonymous referee for their constructive comments. We thank A.~Beloborodov, Z.~Bosnjak, R.~Mochkovitch, S.~Xiong, and B.~B.~Zhang for their comments and suggestions.
Part of the funding for GROND (both hardware as well as personnel) was generously granted from the Leibniz-Prize to Prof. G. Hasinger (DFG grant HA 1850/28-1). This work made use of data supplied by the UK Swift Science Data Centre at the University of Leicester. We thank Taka Sakamoto and Scott D. Barthelmy for the public BAT data.
HFY acknowledges support by the DFG cluster of excellence ``Origin and Structure of the Universe''.
SS acknowledges support by the Th\"uringer Ministerium f\"ur Bildung, Wissenschaft und Kultur under FKZ 12010-514.
PS acknowledges support by DFG grant SA 2001/1-1. 
PS and MT acknowledge support through the Sofja Kovalevskaja Award from the Alexander von Humboldt Foundation of Germany.
SKl and AGN acknowledge support by DFG grant Kl 766/16-1. 
AR, AGN, and AK are grateful for travel funding support through MPE. 
AR acknowledges support by the Th\"uringer Landessternwarte Tautenburg.
KV acknowledges support by DFG grant SA 2001/2-1.
Swift is supported at PSU by NASA grant NAS5-00136.
SKo acknowledges support from the STFC. 
MS is supported by NASA contract NAS5-00136.
TK acknowledges support by the European Commission under the Marie Curie Intra-European Fellowship Programme in FP7. The Dark Cosmology Centre is funded by the Danish National Research Foundation.
\end{acknowledgements}

  \bibliographystyle{aa}
  \bibliography{jbib}

\begin{appendix}
\section{Light curve tables}
\label{sec:Appendix}

\begin{table*}
  \begin{center}
    \caption{Optical reference stars.}
      \begin{tabular}{l c c c c c c} 
        \hline
        R.A. & Dec. & $g'$ & $r'$ & $i'$ & $z$ \\
        \hline
        (J2000) & (J2000) & $\rm mag_{AB}$ & $\rm mag_{AB}$ & $\rm mag_{AB}$ & $\rm mag_{AB}$ \\
        \hline
        10:14:45.18 & -62:21:13.9 & $18.84\pm0.01$ & $18.43\pm0.01$ & $18.30\pm0.02$ & $18.22\pm0.02$ \\
        10:14:48.66 & -62:20:20.5 & $16.77\pm0.01$ & $16.22\pm0.01$ & $16.06\pm0.01$ & $15.94\pm0.01$ \\
        10:14:49.68 & -62:20:29.4 & $17.29\pm0.01$ & $16.16\pm0.01$ & $15.78\pm0.01$ & $15.51\pm0.01$ \\
        10:14:57.84 & -62:21:22.5 & $17.94\pm0.01$ & $17.40\pm0.01$ & $17.24\pm0.01$ & $17.11\pm0.01$ \\
        10:15:03.31 & -62:20:42.4 & $17.82\pm0.01$ & $17.14\pm0.01$ & $16.94\pm0.01$ & $16.82\pm0.01$ \\
        \hline\hline
      \end{tabular} 
      \label{tab:lc_refstars}
  \end{center}
\end{table*}

\begin{table*}
  \begin{center}
    \caption{GROND photometric data $g' r' i' z'$.}
    \begin{tabular}{l l l l l l}
    \hline
    $T_{\rm mid}-T_{0}$ & Exposure & $g'$ & $r'$ & $i'$ & $z'$ \\
    \hline
    s & s & $\rm mag_{AB}$ & $\rm mag_{AB}$ & $\rm mag_{AB}$ & $\rm mag_{AB}$ \\
    \hline
    1120 & 33 & $20.41\pm0.04$ & $19.49\pm0.04$ & $19.12\pm0.04$ & $18.83\pm0.04$ \\
    1229 & 33 & $20.52\pm0.06$ & $19.49\pm0.05$ & $19.16\pm0.06$ & $18.79\pm0.05$ \\
    1338 & 33 & $20.59\pm0.04$ & $19.66\pm0.04$ & $19.17\pm0.04$ & $18.99\pm0.05$ \\
    1447 & 33 & $20.69\pm0.07$ & $19.59\pm0.04$ & $19.20\pm0.05$ & $18.94\pm0.05$ \\
    1592 & 58 & $20.11\pm0.04$ & $19.16\pm0.02$ & $18.78\pm0.03$ & $18.48\pm0.04$ \\
    1787 & 58 & $19.99\pm0.03$ & $18.96\pm0.01$ & $18.59\pm0.03$ & $18.28\pm0.03$ \\
    1981 & 58 & $19.94\pm0.02$ & $19.00\pm0.02$ & $18.59\pm0.03$ & $18.33\pm0.03$ \\
    2173 & 58 & $19.96\pm0.04$ & $19.02\pm0.02$ & $18.61\pm0.03$ & $18.39\pm0.04$ \\
    2374 & 58 & $20.06\pm0.04$ & $19.08\pm0.02$ & $18.69\pm0.03$ & $18.40\pm0.03$ \\
    2568 & 58 & $20.13\pm0.04$ & $19.16\pm0.02$ & $18.76\pm0.03$ & $18.52\pm0.03$ \\
    2760 & 58 & $20.21\pm0.03$ & $19.23\pm0.02$ & $18.85\pm0.03$ & $18.55\pm0.03$ \\
    2953 & 58 & $20.20\pm0.03$ & $19.27\pm0.02$ & $18.89\pm0.03$ & $18.63\pm0.03$ \\
    3154 & 58 & $20.28\pm0.03$ & $19.29\pm0.02$ & $18.90\pm0.04$ & $18.66\pm0.03$ \\
    3346 & 58 & $20.30\pm0.03$ & $19.36\pm0.02$ & $18.99\pm0.03$ & $18.69\pm0.03$ \\
    3539 & 58 & $20.38\pm0.03$ & $19.44\pm0.02$ & $19.02\pm0.03$ & $18.78\pm0.03$ \\
    3737 & 58 & $20.43\pm0.04$ & $19.46\pm0.02$ & $19.12\pm0.03$ & $18.83\pm0.03$ \\
    3940 & 58 & $20.44\pm0.05$ & $19.45\pm0.03$ & $19.08\pm0.04$ & $18.84\pm0.04$ \\
    4126 & 58 & $20.48\pm0.05$ & $19.55\pm0.03$ & $19.13\pm0.04$ & $18.91\pm0.04$ \\
    4320 & 58 & $20.54\pm0.07$ & $19.58\pm0.02$ & $19.16\pm0.03$ & $18.92\pm0.04$ \\
    4518 & 58 & $20.63\pm0.07$ & $19.62\pm0.03$ & $19.28\pm0.03$ & $19.03\pm0.03$ \\
    4736 & 18 & $20.68\pm0.15$ & $19.71\pm0.05$ & $19.35\pm0.06$ & $18.98\pm0.07$ \\
    4736 & 18 & $20.73\pm0.11$ & $19.70\pm0.04$ & $19.35\pm0.05$ & $19.08\pm0.06$ \\
    74739 & 865 & $22.93\pm0.08$ & $21.95\pm0.05$ & $21.59\pm0.06$ & $21.51\pm0.09$ \\
    86363 & 865 & $23.20\pm0.12$ & $21.95\pm0.05$ & $21.68\pm0.06$ & $21.68\pm0.10$ \\
    88178 & 870 & $23.21\pm0.09$ & $22.14\pm0.04$ & $21.80\pm0.07$ & $21.64\pm0.08$ \\
    173620 & 1549 & $>24.46$ & $22.82\pm0.09$ & $22.30\pm0.13$ & $22.61\pm0.19$ \\
    347165 & 2688 & $>24.93$ & $23.52\pm0.12$ & $23.42\pm0.18$ & $23.65\pm0.30$ \\
    1377575 & 2666 & $>24.09$ & $>24.07$ & $>23.69$ & $>23.57$ \\
    1814887 & 4019 & $>25.19$ & $>24.90$ & $>24.16$ & $>23.86$ \\
    \hline\hline
    \end{tabular}
    \tablefoot{All magnitudes have not been corrected for Galactic foreground reddening.}
   \label{tab:lc_optical}
 \end{center}
\end{table*}

\begin{table*}
  \begin{center}
    \caption{GROND photometric data $JHK_{s}$.}
    \begin{tabular}{l l l l l}

    \hline
    $T_{\rm mid}-T_{0}$ & Exposure & $J$ & $H$ & $K_{s}$ \\
    \hline
    s & s & $\rm mag_{AB}$ & $\rm mag_{AB}$ & $\rm mag_{AB}$ \\
    \hline
    301 & 41 & $17.02\pm0.08$ & $17.21\pm0.11$ & $16.85\pm0.12$ \\
    401 & 41 & $17.47\pm0.12$ & $17.26\pm0.13$ & $16.85\pm0.13$ \\
    509 & 41 & $17.37\pm0.12$ & $17.21\pm0.16$ & $16.70\pm0.11$ \\
    619 & 41 & $17.52\pm0.13$ & $17.04\pm0.09$ & $17.14\pm0.14$ \\
    744 & 41 & $17.40\pm0.09$ & $16.98\pm0.06$ & $16.95\pm0.10$ \\
    843 & 41 & $17.76\pm0.10$ & $17.53\pm0.08$ & $17.06\pm0.10$ \\
    927 & 19 & $17.99\pm0.17$ & $17.70\pm0.15$ & $17.74\pm0.28$ \\
    1291 & 205 & $18.36\pm0.09$ & $17.92\pm0.06$ & $17.89\pm0.12$ \\
    1909 & 374 & $17.75\pm0.04$ & $17.51\pm0.04$ & $17.23\pm0.06$ \\ 
    2689 & 373 & $17.96\pm0.05$ & $17.64\pm0.04$ & $17.55\pm0.07$ \\
    3470 & 375 & $18.20\pm0.06$ & $17.95\pm0.05$ & $17.87\pm0.10$ \\
    4254 & 372 & $18.30\pm0.08$ & $17.95\pm0.06$ & $17.95\pm0.10$ \\
    4927 & 238 & $18.49\pm0.13$ & $18.26\pm0.09$ & $18.10\pm0.10$ \\
    5732 & 331 & $18.48\pm0.08$ & $18.46\pm0.08$ & $18.41\pm0.15$ \\
    6577 & 329 & $18.46\pm0.14$ & $18.40\pm0.08$ & $18.70\pm0.16$ \\
    7181 & 190 & $....$ & $18.43\pm0.23$ & $18.98\pm0.28$ \\
    83117 & 7317 & $>20.97$ & $>20.52$ & $>20.14$ \\
    173757 & 1360 & $>20.77$ & $>20.49$ & $>20.15$ \\
    347192 & 2423 & $>20.99$ & $>20.72$ & $>20.28$ \\
    1377597 & 2252 & $>21.00$ & $>20.86$ & $>20.40$ \\
    1814914 & 3607 & $>21.11$ & $>20.84$ & $>20.55$ \\
    \hline\hline

    \end{tabular}
   \tablefoot{All magnitudes have not been corrected for Galactic foreground reddening.}
   \label{tab:lc_nir}
  \end{center}
\end{table*}

\begin{figure*}
  \begin{center}
    \includegraphics[width=10cm]{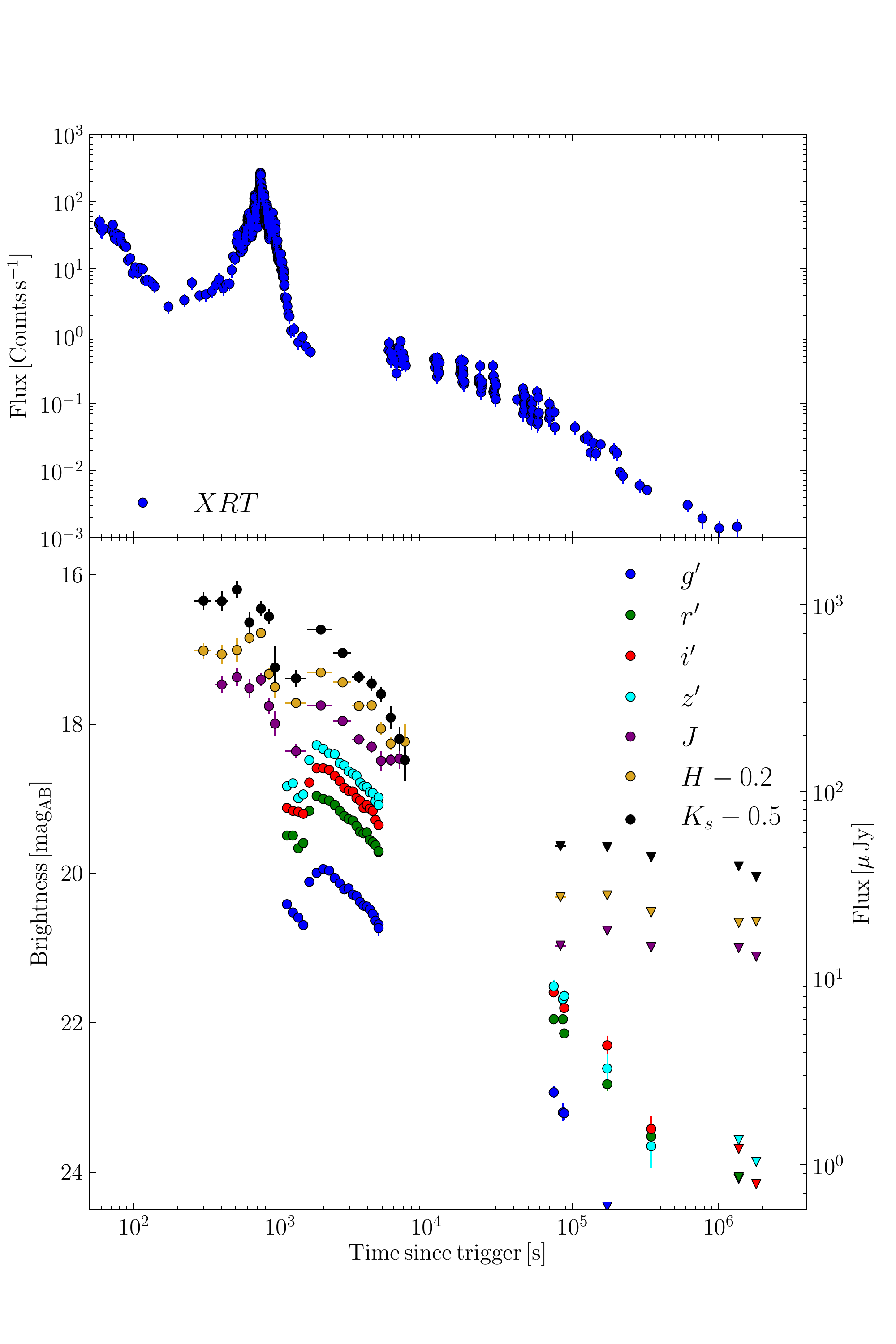}
    \caption{The complete GROND and XRT light curves}
    \label{fig:lc_full}
  \end{center}
\end{figure*}

\end{appendix}

\end{document}